\documentclass[twocolumn]{aastex631}
\usepackage{amsmath}
\usepackage[mathlines]{lineno}\usepackage{ulem}
\usepackage{xcolor}
\definecolor{forestgreen(web)}{rgb}{0.13, 0.55, 0.13}

\def \p {\partial}

\def \. {\cdot}
\def \x {\times}

\begin{document}

\title{ Generation of Surface Sausage Oscillations of a Current Sheet and Propagating Magnetoacoustic Waves by Impulsive Reconnection}
\author{Sripan Mondal}
\affiliation{Department of Physics, Indian Institute of Technology (BHU), Varanasi-221005, India}
\author{A. K. Srivastava}
\affiliation{Department of Physics, Indian Institute of Technology (BHU), Varanasi-221005, India}
\author{David I. Pontin}
\affiliation{School of Information and Physical Sciences, University of Newcastle, Australia}
\author{Eric R. Priest}
\affiliation{Mathematics Institute, St Andrews University, KY16 9SS, St Andrews, UK}




\begin{abstract}
Magnetic reconnection and Magnetohydrodynamic (MHD) waves may well be both playing a role in coronal heating. In this paper, we simulate reconnection in the corona as a response to the convergence of opposite-polarity magnetic sources at the base of the corona. A current sheet forms at a magnetic null and undergoes impulsive bursty reconnection which drives natural modes of oscillation of the current sheet by a process of symbiosis. These are leaky surface sausage modes which cause the length of the current sheet to oscillate. Interaction of the oscillations and reconnection outflows with the magnetic Y-points at the ends of the sheet acts as  sources for magnetoacoustic waves. Fast-mode waves propagate outwards into the coronal environment, while slow-mode waves propagate along the separatrices extending from the ends of the current sheet. The periodicities for sausage oscillations of the current sheet, for the current sheet length, and for the propagating large-scale magnetoacoustic waves are all estimated to be approximately 91 s for the parameters of our experiment.
\end{abstract}


\section{Introduction}
Various coronal observations suggest an \textit{in situ} generation of fast magnetohydrodynamic (MHD) waves in the solar corona during eruptions of flux ropes and filaments, the onset of flares, and the launch of CMEs \citep[e.g.,][and references therein]{1999Sci...285..862N,2013SoPh..288..585S,2014A&A...569A..12N,2018ApJ...853....1S,2020ApJ...894..139Z,2021SoPh..296..169Z,2022ApJ...936L..12W,2024ApJ...962...42H}. Quasi-periodic pulsations in flare light curves are often inferred to be associated with fast MHD waves \citep[e.g.,][and references therein]{2011ApJ...736L..13L,2012ApJ...753...53S,2013A&A...554A.144Y,2017ApJ...844..149K}. \citet{2018ApJ...868L..33L} gave a clear example of the generation of fast MHD waves by magnetic reconnection between coronal loops. Several other examples of the generation and propagation of fast magnetoacoustic wavefronts in the large-scale solar corona and their interaction with various magnetic structures have been reported \citep[e.g.,][and references therein]{2012ApJ...753...52L,2013SoPh..288..585S,2013ApJ...776...58N}, but their physical origin was not clearly explained. Later, other sources of such waves were proposed such as an unwinding jet  \citep{2024ApJ...974L...3Z} or recurrent jets \citep{2018ApJ...861..105S}. \citet{2015A&A...581A..78Z} has also reported the combined presence of the slow and fast magnetoacoustic waves in coronal magnetic structures, while \citet{2023ApJ...953...84M} observed the generation and propagation of quasi-periodic plasma flows in a fan-spine topology and suggested its origin due to periodic reconnection.

The formation of a current sheet (CS) is essential for the occurrence of magnetic reconnection \citep{2014masu.book.....P,2022LRSP...19....1P}. 
Recently, many examples have been proposed for a  symbiosis between waves and reconnection \citep{Sri24,2024ApJ...977..235M,2025ApJ...984...36S}.
For example, \citet{2007PhPl...14l2905L} provided a theoretical model for the transportation of energy from a reconnecting CS due to the propagation of fast MHD waves. More recently, the generation of fast waves produced by coalescence of plasmoids and their interaction with magnetic Y-points (or flare loop-top) in numerical simulations has been reported \citep[e.g.,][and references therein]{2012A&A...546A..49J,2015ApJ...800..111Y,2016ApJ...823..150T,2024ApJ...977..235M}. \Citet{2025ApJ...984...36S} emphasized the formation of CS, reconnection, and heating in the localized solar corona when fast magnetoacoustic waves impinge and pass through a null patch. Also, \citet{2024ApJ...977..235M} suggested that fast waves generated due to plasmoid coalescence can transport sufficient wave energy to large distances to heat the quiet-Sun and coronal holes, as well as accelerate the solar wind above active regions. They reported that the wave properties and their {\it in-situ} coexistence in localized corona may depend on the changing physical conditions of the reconnecting CS in the vicinity. Both these example, demonstrate the two sides of the symbiosis of waves and reconnection \citep{Sri24}. Owing to the second example as depicted above, the mechanisms for the generation of such waves need to be explored in greater detail to understand their effects on coronal heating and mass transport.  

A CS can support a wide variety of MHD waves \citep{Goossens03, Nakariakov05a,Goedbloed10,2019mwsa.book.....R}. For our purpose here it is of interest, in particular, to summarise the main results about surface sausage modes in a one-dimensional CS having an equilibrium magnetic field $B(x)\hat{\bf y}$ and pressure $p(x)$. The modes may either be standing or propagating along the CS in the direction $y$. A surface mode is one whose amplitude decays away spatially in the direction $x$ perpendicular to the sheet, while a sausage mode is one whose $x$-component of velocity is odd in $x$ and so vanishes at the centre ($x=0$) of the sheet. The waves can be leaky and lose energy to the external medium as outward-propagating fast-mode waves when the wavelength is longer than the length of the CS \citep{ 1986SoPh..103..277C, 2014A&A...567A..24H}.

Waves in discrete slabs were first categorised by \citet{1982SoPh...76..239E} using the equations for MHD waves in a one-dimensional medium \citep{1971Phy....53..412G}.  This was then applied by \citet{1986GeoRL..13..373E} to a simple model of a discrete CS of width $2a$, within which the magnetic field vanishes and the plasma pressure ($p_S$) and density ($\rho_S$) are uniform, surrounded by uniform media with plasma pressure $p_e$,  density ($\rho_e$) and magnetic field $\pm B_e$ for $x>a$ and $x<-a$, respectively.  They found two surface sausage solutions having wavenumber $k$ along the sheet, namely, an incompressible Alfv\'en wave with phase speed equal to the external Alfv\'en speed ($v_{Ae}=B_e/(\mu \rho_e)$) and a magnetoacoustic wave whose phase speed when $ka \ll 1$ is equal to
\begin{equation}
\frac{\omega^2}{k^2}=\frac{\rho_e }{\rho_S} ka\ v_{Ae}^2,
    \label{surfacewave}
\end{equation}
which tends to zero as $ka$ tends to zero.  Later, \citet{1997A&A...327..377S} improved the model by allowing the magnetic field to vary continuously with $x$. In this case the Alfv\'en mode is no longer of interest to us since  each field line just oscillates at its own local Alfv\'en speed in the $y$-direction, and so it is not a collective mode. However, the magnetoacoustic surface sausage mode now has a non-zero phase speed when $ka \ll 1$, namely, the maximum value of the tube speed 
\begin{equation}
c_T(x)=\frac{c_S(x)v_A(x)}{(c_S(x)^2+v_A(x)^2)^{1/2}}, 
\label{tubespeed}
\end{equation}
where $c_{S}(x)$ is the sound speed. For example, for the profile $B(x)=B_{e}\tanh(x/a)$, the maximum tube speed is about $0.5 v_{Ae}$ or half the distant external Alfv\'en speed.  

Later, the propagation of waves along a CS was studied using numerical simulations \citep[e.g.,][]{2012A&A...546A..49J,2014ApJ...788...44M}. Also, \citet{2013A&A...550A...1K} interpreted the generation of radio bursts as a manifestation of sausage waves propagating along a CS. In addition, the possibility of explaining the quasi-periodic modulation of flare emission in terms of CS oscillations was discussed by \citet{2009SSRv..149..119N}, while sausage oscillations were detected in cool flare loops \citep{2008MNRAS.388.1899S} and hot EUV loops \citep{2012ApJ...755..113S}.

The goal of this paper is to explore in detail the physical processes at play during driven, time-dependent reconnection that can generate coupled slow and fast magnetoacoustic waves. The following detailed questions are addressed:
\newline
[I] How does the collapse and displacement of a magnetic null to form a CS depend on the imposed converging plasma flows?
\newline
[II] Will the CS  undergo magnetic reconnection and/or intrinsic sausage oscillations?
\newline
[III] If reconnection and/or oscillation of the CS occur, can they generate MHD waves that propagate out into the surrounding medium?
\newline
[IV] What are the physical properties of coupled slow and fast-magnetoacoustic waves propagating in the large-scale corona, and how do their physical properties depend upon the CS dynamics?
\newline

We discuss the numerical setup and methods in Section 2, the detailed  analysis and  results are elaborated in the subsections of Section 3, followed by  concluding remarks in Section 4. 

\section{Setup and Methods Used in the Numerical Experiment} 
We solve the following non-ideal MHD equations to simulate the plasma dynamics in the modeled resistive, thermally conductive and viscous solar corona \citep{2014masu.book.....P,2017ApJ...841..106Z,2024ApJ...977..235M}:
 \begin{equation} 
\frac{\partial \rho}{\partial t} + \vec{\nabla} \cdot ( \rho \vec{V} ) = 0,
\end{equation}
\begin{equation}
  \frac{\partial}{\partial t}(\rho \vec{V}) + \vec{\nabla} \cdot \left [ \rho \vec{V}\vec{V}  + P_{tot}\vec{I} - \frac{\vec{B}\vec{B}}{4\pi} \right ] = \mu \vec{\nabla} \cdot [2S-\textstyle{\frac{2}{3}}(\vec{\nabla} \cdot \vec{V})\vec{I}] ,
\end{equation}

\begin{equation}
\begin{split}
\frac{\partial e}{\partial t} +  \vec{\nabla} \cdot \left( e\vec{V} + P_{tot}\vec{V} -\frac{\vec{B}\vec{B}}{4\pi} \cdot \vec{V}\right)  = \eta \vec{J^{2}}-
   \vec{B} \cdot \vec{\nabla} \times (\eta \vec{J})\\
   +\vec{\nabla}_{\parallel} \cdot (\kappa_{\parallel} \vec{\nabla}_{\parallel} T)+\mu [2S^{2}-\textstyle{\frac{2}{3}}(\vec{\nabla} \cdot \vec{V})^{2}],
\end{split}   
\end{equation}

\begin{equation}
  \frac{\partial \vec{B}}{\partial t} + \vec{\nabla} \cdot \left(\vec{V}\vec{B} - \vec{B}\vec{V}\right)+ \vec{\nabla} \times (\eta \vec{J}) = 0,
\end{equation}

\quad \textrm{where} \quad
\begin{equation}
P_{tot} = P + \frac{B^2}{8\pi}, ~~e = \frac{P}{\gamma-1} + {\textstyle{\frac{1}{2}}\rho V^{2}} + \frac{B^2}{8\pi}
\end{equation}
\quad \textrm{and} \quad
\begin{equation}
  \vec{J} = \frac{\vec{\nabla} \times \vec{B}}{4\pi}, ~~\vec{\nabla} \cdot \vec{B} =0.
\end{equation} 
The field-aligned thermal conduction tensor is taken to be $\kappa_{\parallel}=10^{-6}~T^{5/2}~\mathrm{erg~cm^{-1}~s^{-1}~K^{-1}}$. The dynamic viscosity coefficient ($\mu$) is $0.027~\mathrm{g~cm^{-1}~s^{-1}}$. The strain rate tensor ($S_{ij}$) is given by $\frac{1}{2}(\partial V_{i}/\partial x_{j}+\partial V_{j}/\partial x_{i})$. A uniform resistivity of $2.4\times 10^{8}~ \mathrm{m^{2}s^{-1}}$ is taken throughout the domain without any kind of localized or current-dependent enhancement which is similar to resistivity typically used in numerous coronal reconnection simulations \citep[e.g.,][]{2019ApJ...872...32S,2022A&A...666A..28S,2023A&A...675A..97F,2024ApJ...963..139M,2024ApJ...977..235M}. A uniform number density of $10^{9}~\mathrm{cm^{-3}}$ and a temperature of 1 MK is assumed to ensure an initially homogeneous solar coronal plasma throughout the domain neglecting gravity. In the solar corona, the scale-height over which the plasma pressure and density decline with height is typically 50 Mm. The effect of gravity is therefore not large in the corona \citep[see, for example,][]{2019ApJ...872...32S}. Therefore, for simplicity, we have neglected the effect of gravity and stratification  in the present simulation while focusing solely on coronal dynamics. 

\begin{figure*}
\hspace{-2 cm}
\includegraphics[width=1.2\textwidth]{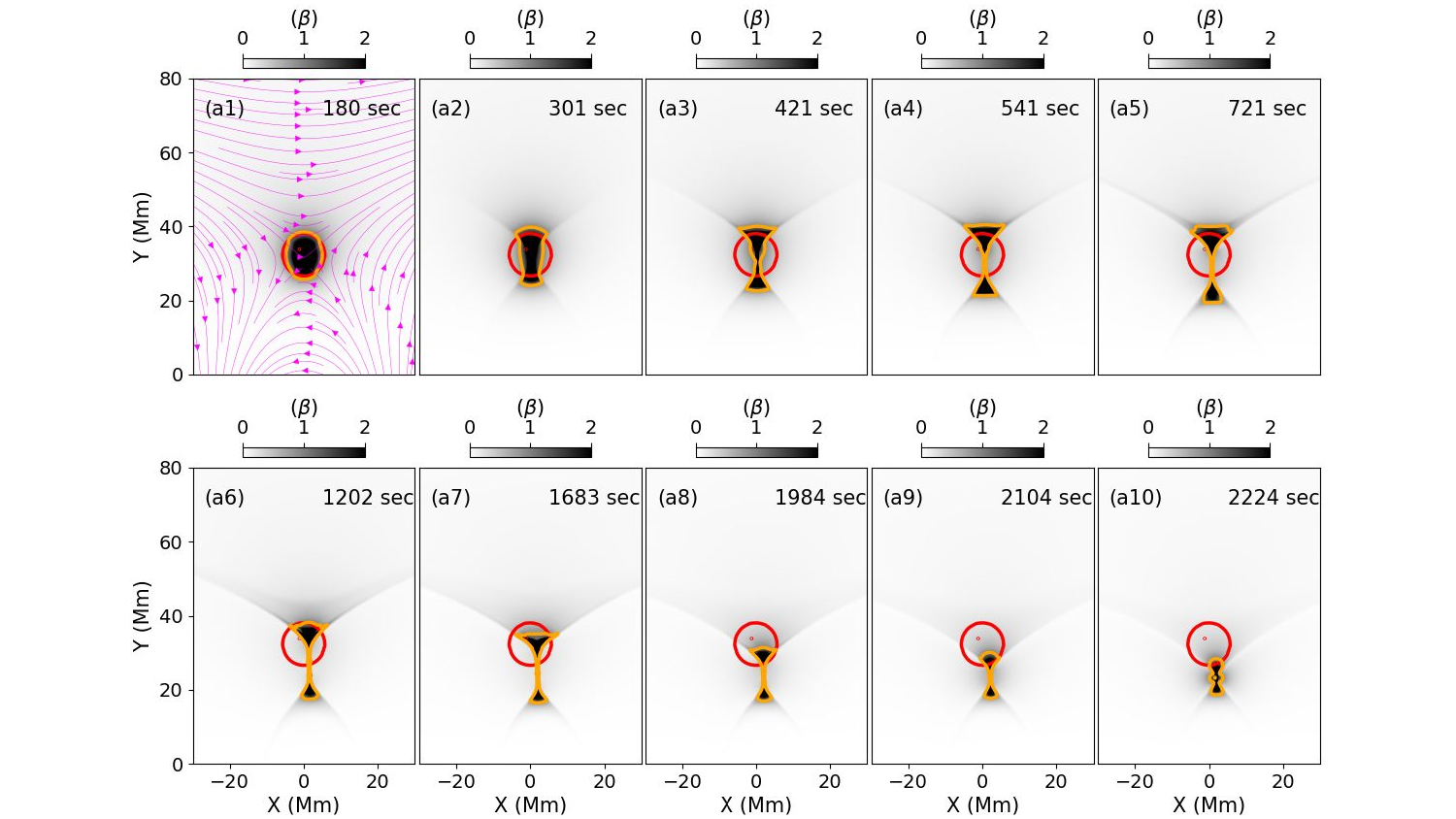}
\caption{\texttt{Collapse and displacement of the null and the high-$\beta$ region surrounding it:} Panels (a1)-(a10) reveal a gradual collapse of the null to form a current sheet-like structure (see panel (a1)-(a6)) followed by a gradual relaxation at later stages  (see panels (a7)-(a10)). Magenta streamlines in panel (a1) depict the magnetic field configuration at $t = 0$ s. The red circle denotes the initial unperturbed plasma $\beta$ = 1 contour in each panel, whereas the orange curve reveals the deformed plasma $\beta$ = 1 contour. It is clear that the current sheet-like structure is displaced rightward and downward  from its initial position. An animation showing the entire evolution and displacement of the null region from 0 to 2224 s is available in the online version. The real time duration of the animation is 6 s.}
\label{label 1}
\end{figure*}

We follow \citet{2019ApJ...872...32S} to construct the initial setup of a potential magnetic field, namely,
\begin{equation}
\vec{B} = B_{0}\hat{x}+ \frac{F\vec{r_{1}}}{\pi r_{1}^{2}}-\frac{F\vec{r_{2}}}{\pi r_{2}^{2}},
\end{equation}
\textrm{where} 
\begin{equation}
\vec{r_{1}}=(x-x_{0})\hat{x}+(y-y_{0})\hat{y}; ~~\vec{r_{2}}=(x+x_{0})\hat{x}+(y-y_{0})\hat{y}.
\end{equation}
Two opposite magnetic fragments with magnetic flux ($F$) of $ \pm 3.6\times 10^{11}~$ G cm are placed 40 Mm (i.e., $x_{0}= 20~\mathrm{Mm}$) apart  at a depth of 20 Mm (i.e., $y_{0}= -20~\mathrm{Mm}$) from the bottom boundary of our two-dimensional  domain, which extends from -80 Mm to +80 Mm in the $x$-direction and 0 to 80 Mm in the $y$-direction. Finally, we add an overlying uniform magnetic field of 15 G ($B_{0}\hat{x}$) which results in the presence of a very localized high plasma-$\beta$ region of radius of approximately 6 Mm containing a magnetic null point at $x= 0~\mathrm{Mm}$ and $y= 31.6~\mathrm{Mm}$ (see Figure \ref{label 1}(a1)). Numerical stability of this initial magnetohydrostatic equilibrium with a current-free magnetic field in a homogeneous plasma is confirmed by a test run.

We impose a converging motion in the $x$-direction by using a velocity perturbation at the bottom boundary given by
\begin{equation}
V_{x}(x,t) = -V_{0}g(x)f(t)
\end{equation}
\quad \textrm{where} \quad
\begin{equation}\label{eq:g(x}
g(x)=\tanh (0.08(x-x_{as})),
\end{equation}
\quad \textrm{and} \quad
\begin{equation}\label{eq:f(t)}
f(t) =
\begin{cases}
\frac{t}{t_1} & \text{when $t \leq t_1$;} \\
1 & \text{when $t_1 \leq t \leq t_2$;} \\
\frac{t(t_3-t)}{t_2(t_3-t_2)} & \text{when $t_2 \leq t \leq t_3$;} \\
0 & \text{when $t> t_3$,}
\end{cases}
\end{equation}
where $t_1=257.6$~s, $t_2=1459.8$~s and $t_3=2146.8$~s. Since the bottom boundary of the simulation domain corresponds to the base of the solar corona, we impose $V_{0} = 10~\mathrm{km\ s^{-1}}$ which is comparable to the converging velocities used in several previous coronal simulations \citep[e.g.,][]{2021A&A...646A.134J,2022A&A...668A..47B,2023A&A...673A.154L,2025A&A...696A.158L}. Since, in general, the footpoints of the polarities may be driven asymmetrically, we mimic such an asymmetry by placing the inversion point of the $tanh$ profile at $x_{as} = 10~\mathrm{Mm}$, while the magnetic polarities are placed symmetrically about $x = 0~\mathrm{Mm}$ (see the Appendix A). Even though we are not using random multiple velocity perturbations having different spatial locations and occurrence times, the proposed spatial asymmetry along with the time-dependent amplitude of the velocity perturbations make the boundary driving general. 

We numerically solve the above coupled partial differential equations (PDEs) (3-6) using open source MPI-AMRVAC 3.0\footnote{https://amrvac.org/} \citep{2023A&A...673A..66K}.  All of the physical quantities in Eqs. 3-13 are subsequently normalized with respect to their typical values in the numerical code, namely, $L^{*}= 10^{9}~\mathrm{cm}$, $\rho^{*}= 2.34\times10^{-15}~\mathrm{g~cm^{-3}}$, $V^{*}= 1.16\times10^{7}~\mathrm{cm~s^{-1}}$, $T^{*}= 10^{6}~\mathrm{K}$, $P^{*}=0.3175~\mathrm{dyne~cm^{-2}}$, and $B^{*}= 2~\mathrm{G}$. Five AMR levels are used which results in an effective number of grid points of 4096 × 2048, with the smallest grid size being 39 km in both directions. A $``$two-step$"$ method is used for performing the temporal integration, while the estimates of the flux at cell interfaces is done using a $``$Harten-Lax-van Leer (HLL)$"$ Riemann solver \citep{1983JCoPh..49..357H}. Spurious numerical oscillations  are suppressed by using a second-order symmetric total variation diminishing limiter, namely $``$vanleer$"$ \citep{1979JCoPh..32..101V}. Except the bottom one, continuous boundary conditions are imposed at other three boundaries to ensure zero gradients for all the variables across them. At the bottom boundary, pressure and density are fixed as per the initial conditions, whereas the $y$-component of the velocity is taken to be anti-symmetric. For the magnetic field, the $x$-component is kept symmetric whereas the $y$-component is fixed to its initial condition.
\begin{figure*}
\hspace{-1.5 cm}
\includegraphics[scale=0.275]{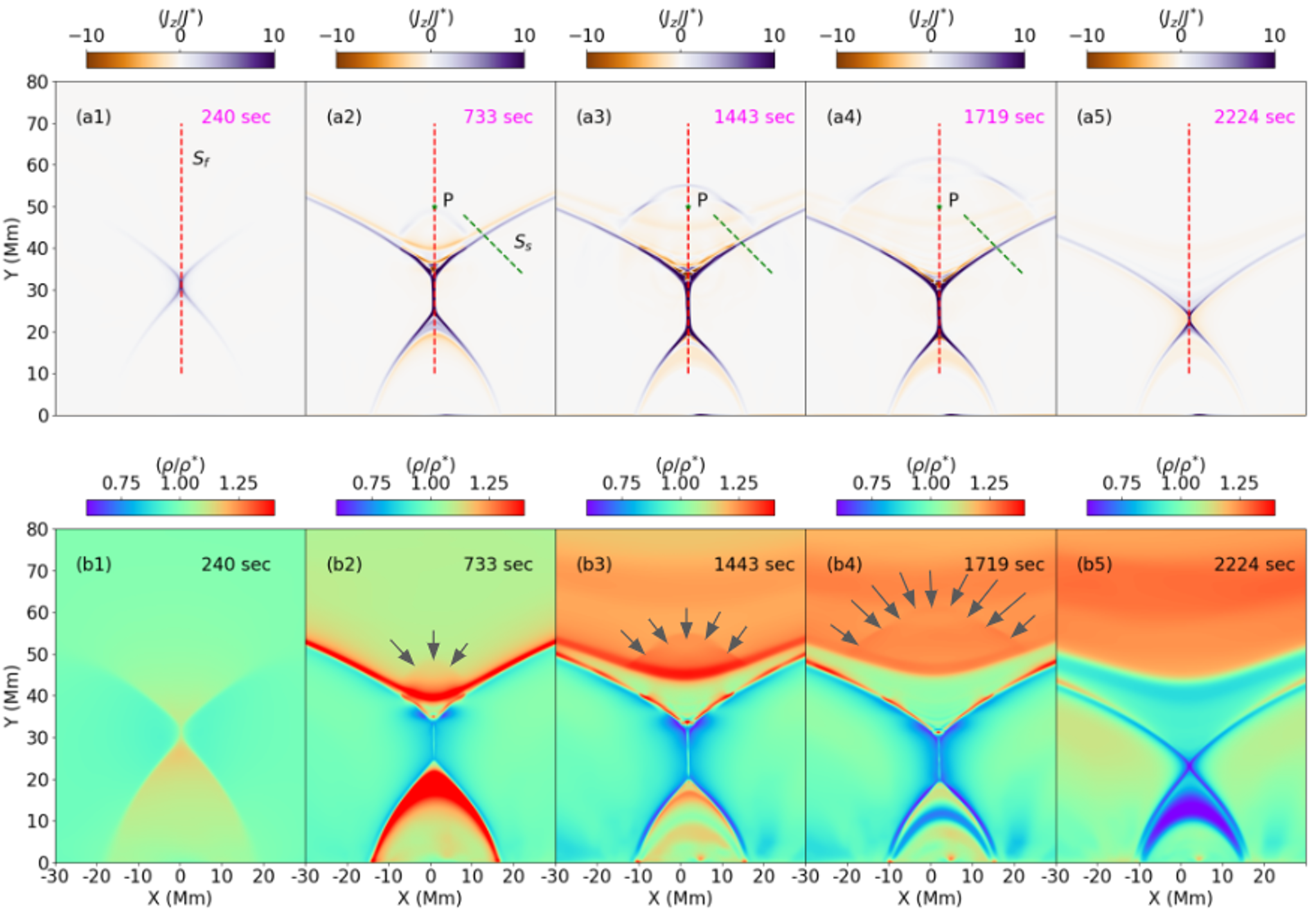}
\caption{\texttt{The temporal evolution of $J_{z}$ and $\rho$:} Panels (a1)-(a5) exhibit the accumulation of current at 240 s, CS and wave-like features at 733 s, 1443 s and 1719 s, and the decay of current at 2224 s. The $y$-directed red dashed line denotes the time-dependent positions of the slit (denoted as $S_{f}$) within $y$ = [10,70] Mm at $x$ = $x_{\mathrm{min(B)}}(t)$ Mm at time $t$. `P' denotes the location at which the identification of the arc-shaped wavefronts is carried out (see panel (a)-(c) of Figure~\ref{label 8}). The tilted green dashed line is used to locate the position of the slit `$S_{s}$' in order to identify the disturbances propagating along the separatrix (see panels (d)-(f) of Figure~\ref{label 8}). Panels (b1)-(b5) reveal the corresponding evolution in density. Panels (b2)-(b4) reveal the existence of both arc-shaped perturbations (depicted via arrows) and disturbances moving along the separatrices as high-density patches. An animation showing the entire dynamics in current density and density from 0 to 2224 s is available in the online version. The real time duration of the animation is 9 s.}
\label{label 2}
\end{figure*}
\section{Results $\&$ Discussions}
Imposed time-varying converging velocity perturbations cause lateral movements of the footpoints of the coronal magnetic field geometry at the base of the model corona. Basically, the local maxima of the opposite polarities start at a separation of around 44 Mm at the base of the model corona which reduces to a minimum value of 32 Mm before the velocity driving starts to diminish at the later stages of the simulation as prescribed by $f(t)$ (see the Appendix A).  Moreover, the spatial variation of the perturbation is such that the velocity gradient across the coronal field lines emanating from the positive polarity is higher than those anchored at the negative polarity. Due to the presence of these gradients, the magnetic fields become compressed with time to a greater or lesser degree at the opposite polarities (see Appendix A for more details). As a result of these movements of the field lines connected to opposite polarities towards each other, magneto-plasma processes such as collapse of the magnetic null, accumulation of current, and formation of the CS take place in a sequential manner. Here, we discuss each process as follows.
\begin{figure*}
\hspace{0.5 cm}
\includegraphics[scale=0.2]{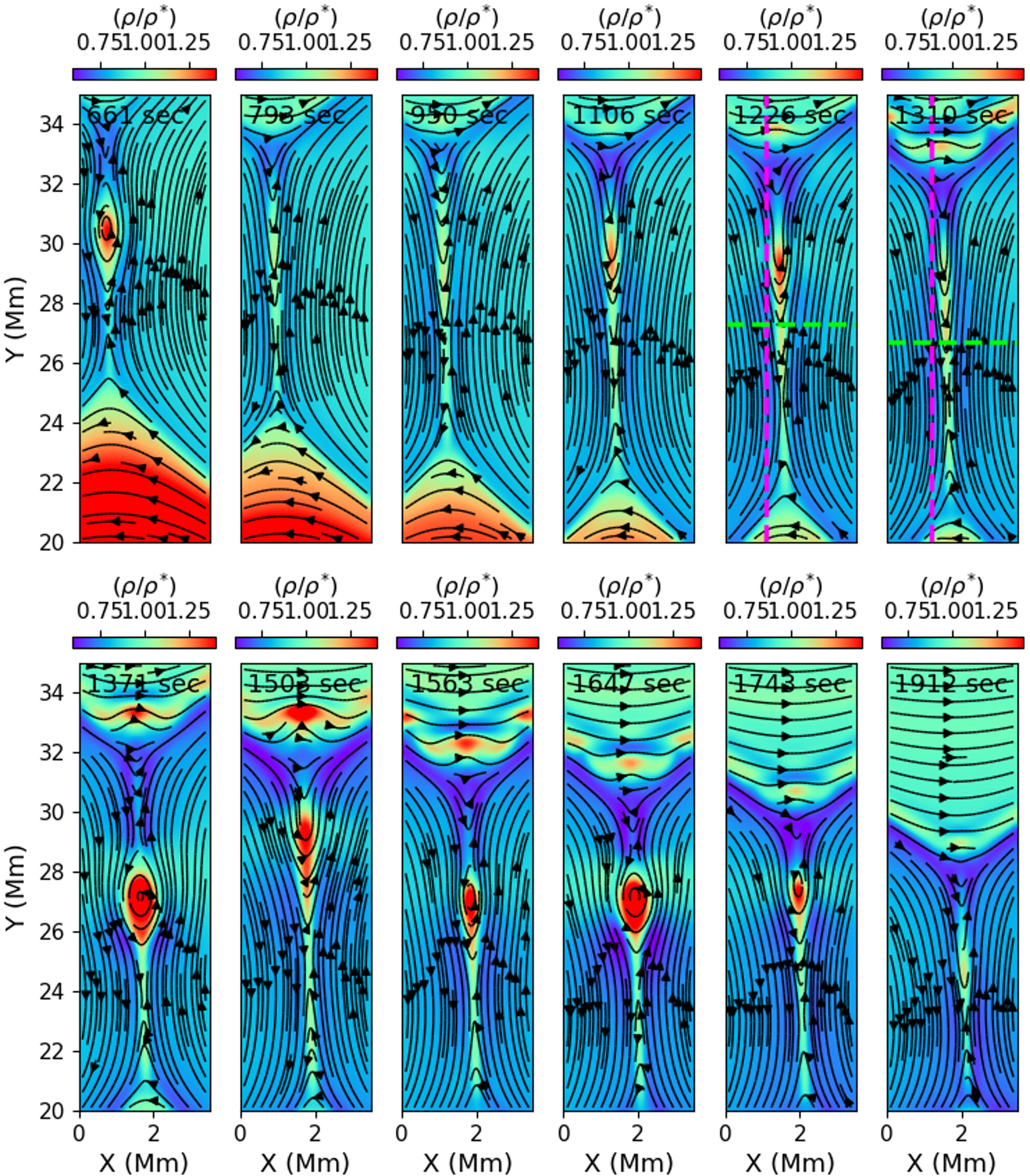}
\caption{\texttt{Identification of plasmoids:} Closed magnetic field lines are overplotted on high-density plasma blobs in density maps of the field of view $x$ = [0, 3.5] Mm and $y$ = [20, 35] Mm, which reveal that plasmoids are formed  at five instances, namely, around 661 s, 1371 s, 1563 s, 1647 s and 1743 s. At other times, even though there are  high-density plasma bulges or flows that form and propagate, they are not plasmoids, since they do not possess closed field lines. Magenta and lime dashed lines in the panels of 1226 s and 1310 s indicate the locations of slits used in extracting the $V_{x}$ profiles shown in the top and bottom panels of Figure~\ref{label 6}, respectively. An animation of real time duration  4 s showing a close view of the distinctive features of plasma bulges and plasmoids from 469 s to 1984 s is available in the online version. }
\label{label 3}
\end{figure*}
\subsection{Collapse and Displacement of the Magnetic Null and its High-$\beta$ Environment}
As the coronal field lines connected to opposite polarities converge towards each other driven by the imposed motions at the base of the corona, the initial high-$\beta$ region (containing a magnetic null) becomes compressed in the $x$-direction and elongated in the $y$-direction (see Figure~\ref{label 1}(a1)-(a6)).  At later times, once the amplitude of the velocity driving decreases in time, the high-$\beta$ region returns towards its equilibrium configuration  (see Figure~\ref{label 1}(a7)-(a10)). The magnetic null and its high-$\beta$ environment not only collapse, but also undergo a simultaneous rightward and downward displacement from their initial location. In Figure~\ref{label 1}, the red circle denotes the initial unperturbed contour of the $\beta$ =1 region, whereas the time-varying orange contour marks out its later position and orientation. In the following sub-sections we describe the local dynamics within this high-$\beta$ region, and the associated wave evolution in the surrounding corona.

\subsection{Morphological Description of the  Resulting Dynamics}
As discussed in the previous subsection, the null and its high-$\beta$ surrounding regions undergo a gradual collapse in time. Such a collapse corresponds to a higher gradient in the inflow magnetic field ($B_{y}$). This process further results in a gradual accumulation of  current at the intersection of the separatrices, and also along them from around 144 s onwards (see the animation associated with Figure~\ref{label 2}). Due to the  accumulated current and magnetic field, an unbalanced Lorentz force starts to accelerate the collapse of the null, and the formation of a CS. The CS first grows in length in the $y$-direction with time (see Figure~\ref{label 2}(a1)-(a4)), and then at a later stage the CS shrinks in  length (see Figure~\ref{label 2}(a5)). Formation and shrinkage of the CS is also  visible in the density maps in Figure~\ref{label 2}(b1) and (b5). In addition, the density maps exhibit the generation and propagation of arc-shaped large-scale wave-like features and high-density patches along the separatrices mostly from the top Y-point (see Figure~\ref{label 2}(b2-b4)). 

Previous studies suggest that such a generation of waves is mostly due to pressure perturbations generated by interaction of plasmoids and/or reconnection outflows with magnetic Y-points. Therefore, we undertake  a closer inspection of the dynamics within the CS itself. Interestingly, we find that relatively high-density plasma bulges move preferentially in the upward direction along the CS due to the higher plasma pressure gradient in that direction from around 529 s to 1984 s, and they interact with the top magnetic Y-point multiple times around 613 s, 685 s, 830 s, 974 s, 1046 s, 1130 s, 1250 s, 1347 s, 1395 s, 1491 s, 1515 s, 1587 s, 1671 s, 1755 s, 1791 s and 1948 s (see the animation associated with Figure~\ref{label 2} for a visual inspection). Such interactions  generate  large-scale arc-shaped wavefronts and high-density patches propagating along the upper separatrices. In addition, such interactions also take place at the bottom Y-point at multiple times, for example, around 1443 s, 1527 s, 1611 s, 1683 s, 1719 s, 1827 s, 1888 s etc as evident from running difference density maps in Appendix B. 

Even though the generation and propagation of strong arc-shaped wavefronts are not visible from the bottom Y-points, the generation of high-density patches and their propagation along the loops below the bottom Y-point are visible (see Appendix B for more details). Therefore, we examine whether all of these plasma bulges can be classified as plasmoids or not. Plasmoids will have closed magnetic flux within them.  We find such closed magnetic field lines only at five instances, i.e., around 661 s, 1371 s, 1563 s, 1647 s and 1743 s (see Figure~\ref{label 3} and its associated animation). These plasmoids eventually collide with the top Y-point around  685 s, 1395 s, 1587 s, 1671 s and 1755 s, respectively. So, plasmoids cannot be  responsible for the generation of all the large-scale wavefronts from the magnetic Y-point in the present simulation. Since  the observed generation of wave-like features during filament eruptions and flares is of interest, we perform a detailed analysis of the CS dynamics to understand the underlying mechanism for the generation of the wavefronts in the following sections. 
\begin{figure*}
\hspace{-1.5 cm}
\includegraphics[scale=0.275]{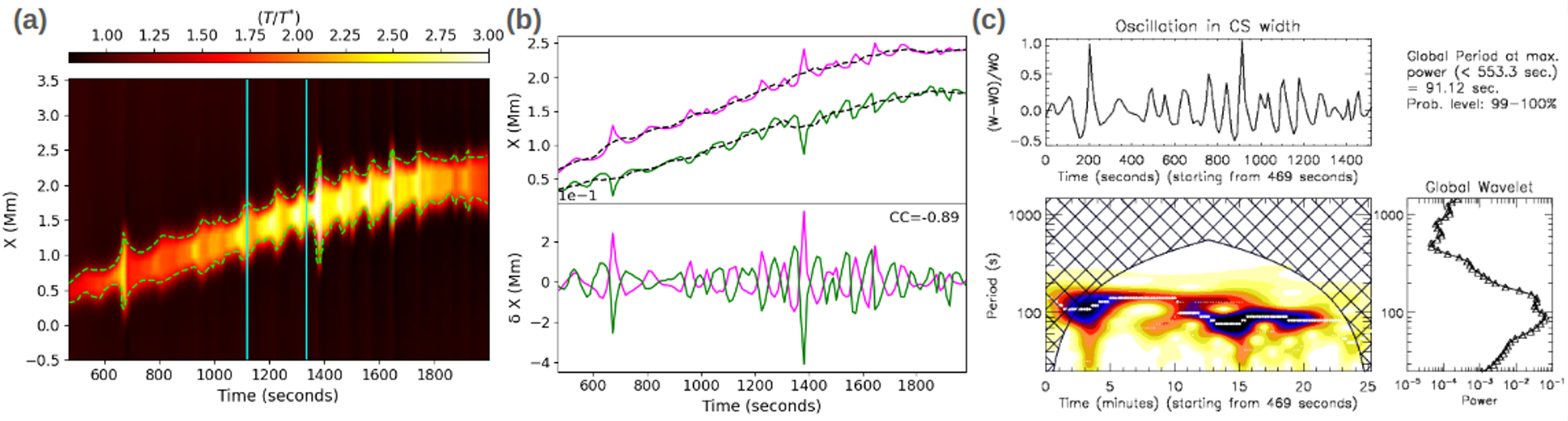}
\caption{\texttt{Sausage oscillation of the CS:} Panel (a) shows the time-distance diagram in temperature estimated across the CS within $x = [-0.5, 3.5]$ Mm at the instantaneous mid-point of the CS. Lime dashed curves denote the repetitive compression and expansion of the CS width, suggestive of sausage oscillation. Cyan solid lines reveal the time window considered for detailed analysis of properties of such oscillations in the absence of plasmoids. Panel (b) exhibits the in-phase relation of the oscillations at the left and right edges of the CS. Dashed black curves in the top panel denote the subtracted background motion of the CS associated with its gradual rightward movement. The difference between the coordinates of these black curves at each time is basically $W_0$. Panel (c) exhibits the wavelet estimate of $(W-W_0)/W_0$ where $W_0$ is the distance between the right and left edges of the CS excluding the localized variations in their mean positions at each time. $W$ is basically the distance between such edges including localized variations of their positions at each time. So, subtracting $W_0$ means subtracting the background trend associated with the slow rightward motion of the CS. Hence, the subtracted profile exhibits a sausage oscillation of the CS with a periodicity of $\approx $91 s with one $\sigma$ uncertainty of about 10 s.}
\label{label 4}
\end{figure*}
\subsection{Oscillatory Expansion and Contraction of the CS Cross-Section: Sausage Oscillation}
The movement of plasmoids and/or plasma bulges along the CS  creates a variation in the cross section of the CS at any particular location along the sheet, such as at the mid-point. Since the CS is becoming elongated in the initial stages and squeezed at later times, its mid-point will not remain at exactly the same location for a long duration. Therefore, we first need to find the instantaneous positions of the top and bottom Y-points so as to be able to determine the instantaneous midpoint of the CS. Since, the CS itself exhibits a rightward motion, estimating the positions of the Y-points first requires determining the $x$-coordinate of the CS axis. So, we consider the distribution of the current along the $y$-direction at the $x$-location at which the magnetic field attains its minimum value. Hereafter, such a time evolving $x$-coordinate is denoted as $x_{\mathrm{min(B)}}(t)$, where $t$ stands for an instantaneous time. Basically, we estimate the location of the bottom Y-point ($\mathrm{Ypoint_{bottom}(t)}$) as $y-\Delta y$ such that $B_{x}[x_\mathrm{{min(B)}}(t),y-\Delta y] > 1~$Gauss and $B_{x}[x_\mathrm{{min(B)}}(t),y] < 1~$Gauss. Similarly, $y$ is considered as the $y$-coordinate of the top Y-point ($\mathrm{Ypoint_{top}}(t)$) such that $B_{x}[x_\mathrm{{min(B)}}(t),y-\Delta y] < 1~$Gauss and $B_{x}[x_\mathrm{{min(B)}}(t),y] > 1~$Gauss, where $\Delta y$ is the grid spacing. Therefore, the midpoint of the CS about its length is estimated using the instantaneous $y$-coordinates of the top and bottom Y-points.  
\begin{figure*}
\hspace{-1.5 cm}
\includegraphics[scale=0.275]{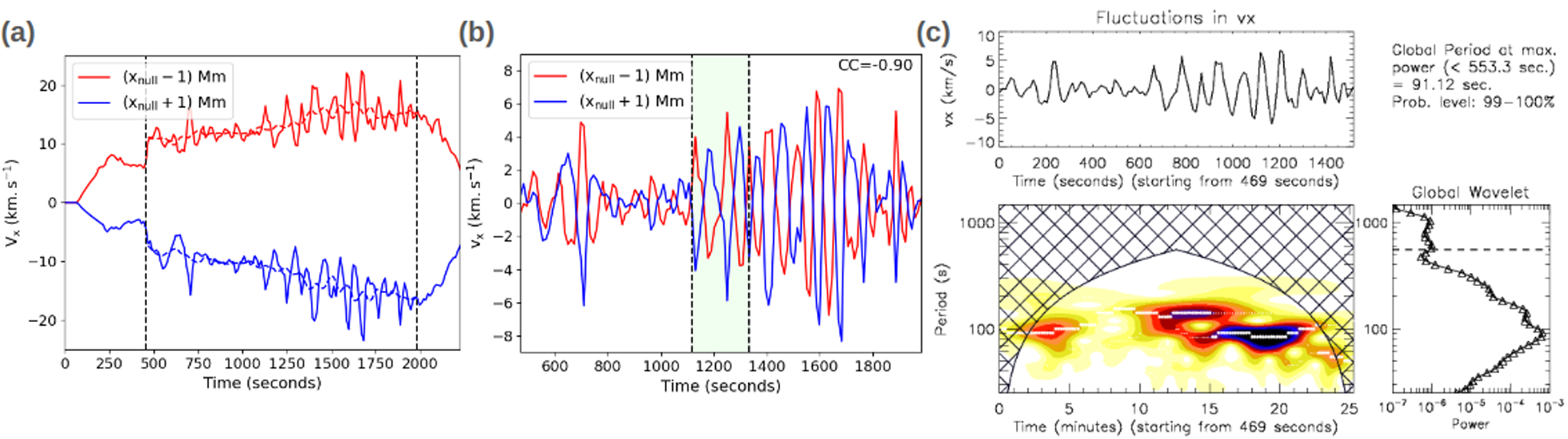}
\caption{\texttt{Periodic and symmetric variation of the velocity component normal to the CS axis}: Panel (a) shows variations in the average $V_{x}$ with time. Averaging is carried out within $y(t)$ = [$\mathrm{Ypoint_{bottom}}(t), \mathrm{Ypoint_{top}}(t)$] Mm at $x$ = ($x_\mathrm{min(B)}(t)-1$) Mm and  $x$ = ($x_\mathrm{min(B)}(t)+1$) Mm (labeled as $(x_\mathrm{null} -1)$ Mm and $(x_\mathrm{null} -1)$ Mm, respectively). Black dashed vertical lines denote the start and end of the sausage oscillations. Red and blue dashed curves from 469 s to 1984 s denote the long-term background trends in averaged $V_{x}$. Panel (b) reveals a strong anti-correlation in background subtracted fluctuations in averaged $V_{x}$ (denoted as $v_{x}$) with a cross correlation coefficient of -0.90. These oscillations in $v_{x}$ at the left and right sides of the CS are basically in-phase with each other with an inherent symmetry about the CS axis. The shaded region reveals the time window within which the further analysis  of the sausage mode is carried out in the absence of plasmoids (the same as the time window within two cyan lines in Figure~\ref{label 4}(a)). Panel (c) exhibits the wavelet estimation of the background subtracted fluctuations in averaged $v_{x}$ as estimated at $x$ = ($x_\mathrm{min(B)}(t)-1$) Mm. Such a fluctuation in average $v_{x}$ possesses a periodicity of $91 \pm 6$ s, consistent with the periodicity associated with the oscillation of the cross-section of the CS.}
\label{label 5}
\end{figure*}

To examine whether there is any oscillation in the cross-section of the CS with time, we take a horizontal slit across the CS from $x = -0.5$ Mm to $x = 3.5$ Mm at the instantaneous midpoint of the CS  from 469 s to 1984 s. The side edges of the CS are found to oscillate, 
resulting in a modulation of its cross-sectional area (see lime dashed curves in Figure~\ref{label 4}(a)). Now, plasmoids are not present frequently within the CS, so such an oscillation in cross section is not  due to the passage of plasmoids alone. Since the high-temperature CS moves to the right on a long time-scale, we subtract the long-term background trend associated with such a motion. Basically, we are interested in tracking the oscillation in the cross-section of the CS, and so such subtraction of the background slow rightward motion (denoted by black dashed curves in the top panel of Figure~\ref{label 4}(b)) is necessary. We estimate a cross-correlation coefficient of $-0.89$ between the background subtracted boundary oscillations on each side of the CS (see bottom panel of Figure~\ref{label 4}(b)). 

Such strong anti-correlation between the background subtracted oscillations on each side of the CS indicates a periodic thinning and fattening of the CS, i.e., the presence of sausage mode oscillations. We estimate the periodicity associated with variations in $(W-W_0)/W_0$ using the unique wavelet and randomization technique \citep{1985AJ.....90.2317L,2001A&A...368.1095O}, where $\mathrm{W}$ stands for the difference between instantaneous locations of the right and left edges of the CS (denoted via magenta and green curves respectively in top panel of Figure~\ref{label 4}(b)), whereas $W_0$ is the difference between the subtracted background coordinates on the right and left edges associated with the slow rightward motion of the CS (denoted via black dashed curves in top panel of Figure~\ref{label 4}(b)). A period of $\approx$ $91 \pm 10$ s is associated with such an oscillation in the CS width, i.e, with a sausage oscillation (see Figure~\ref{label 4}(c)). A  wavelet analysis performed directly on $W$ gives the same periodicity together with extra power at very low frequencies due to $W_0$. Moreover, since the oscillation in the length of the CS may cause the midpoint to be located at different positions along the CS at different times, we have verified that the  oscillation in width is not an artefact caused by the length oscillation by recomputing it at two fixed heights and finding the same periodicity. Now, since the presence of sausage modes will perturb the velocity component ($V_{x}$) normal to the CS axis,  let us try to understand more detailed properties of the sausage mode in terms of $V_{x}$ in following subsections.
\begin{figure*}
\hspace{-1.5 cm}
\includegraphics[scale=0.825]{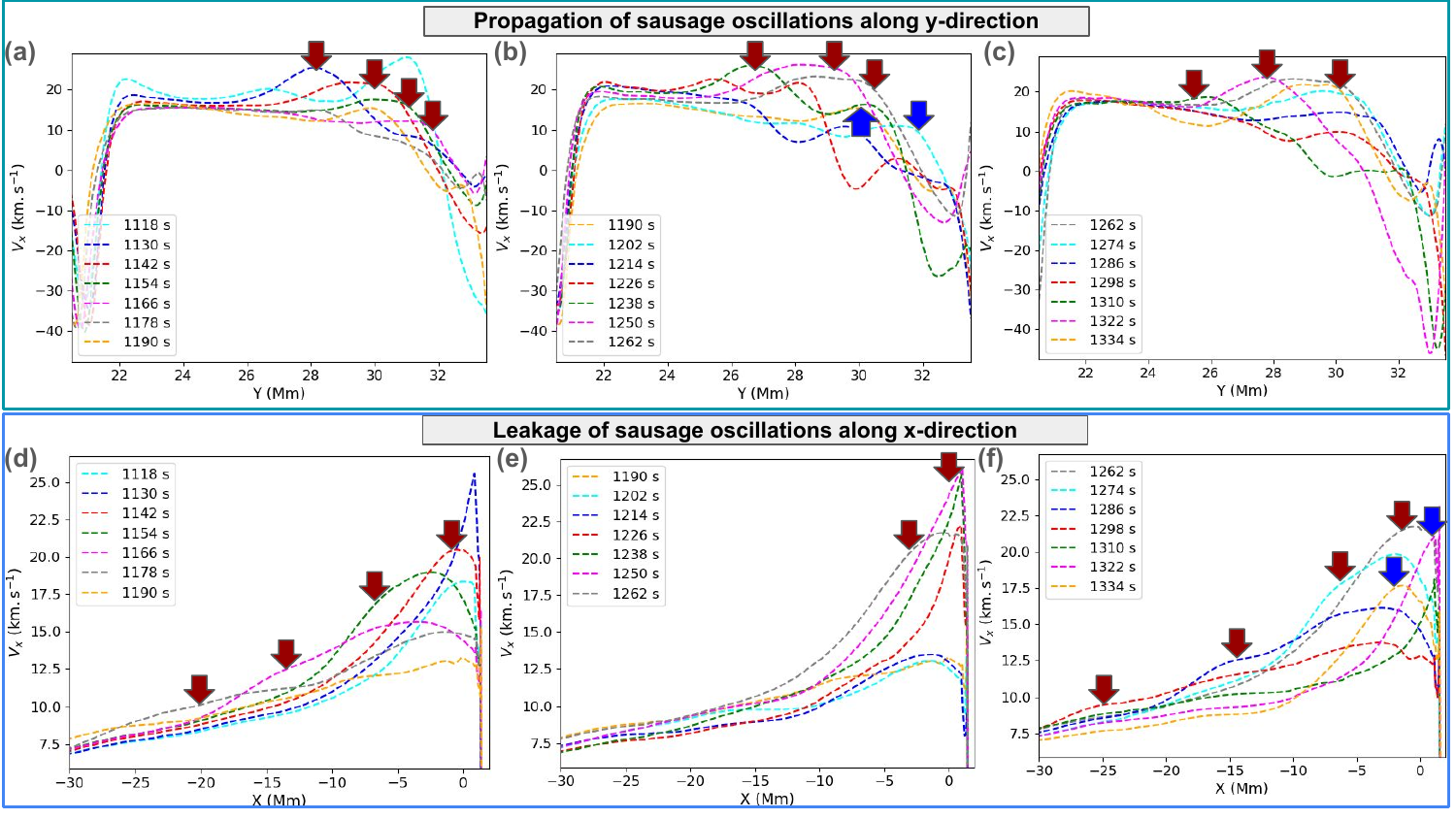}
\caption{\texttt{Characteristics of the Sausage Mode: Propagating and Leaking:} Panels (a), (b) and (c) exhibit successive outward propagation of wavy profiles in $V_{x}$ along the CS in the $y$-direction as depicted by arrows from 1118 s to 1334 s in the absence of plasmoids, confirming the propagating nature of the sausage mode. These profiles of $V_{x}$ are obtained at the left edge of the CS, i.e., 0.5 Mm away from the axis of the CS when the CS half-width is about 0.15-0.2 Mm. Panels (d), (e) and (f) reveal that $V_{x}$ decreases in the $x$-direction at distances away from the CS. On top of such overall trends, wavy features are found to propagate outward in the $x$-direction as denoted by arrows which  gradually diminish in amplitude close to the CS. Such a scenario  suggests the sausage mode has a leaky nature.}
\label{label 6}
\end{figure*}
\subsubsection{Temporal Variation of the Average Reconnection Inflow Velocity}

We estimate the average $V_{x}$ (reconnection inflow velocity) between $\mathrm{Ypoint_{top}(t)}$ and $\mathrm{Ypoint_{bottom}(t)}$ along $x_{\mathrm{min(B)}}(t)\pm 1$~Mm. Until 96 s, there is no $V_{x}$. Beyond that it increases almost linearly up to $\mathrm{9~km\ s^{-1}}$ until around 250 s and thereafter saturates at slightly lower values (see Figure~\ref{label 5}(a)). After 469 s, $V_{x}$ grows and oscillates with time until 1984 s. Beyond 1984 s, $V_{x}$  drops rapidly with time. Here, we are specifically interested in the oscillations in $V_{x}$ which could possibly represent sausage mode oscillations. To disentangle such oscillations from the long-term background variation, we subtract the long-term variations depicted as red and blue dashed curves in Figure~\ref{label 5}(a). From Figure~\ref{label 5}(b), it is evident that the background subtracted fluctuations in $V_{x}$ (denoted as $v_{x}$) at the left and right edges of the CS are symmetric with each other with a cross-correlation coefficient of 0.90. We find the periodicity is about $91 \pm 6$ s for the fluctuations at the left edge of the CS (see Figure~\ref{label 5}(c)). This is the same as the periodicity associated with the variation in the cross section of the CS. This confirms the presence of sausage oscillations. 

\subsubsection{Identification of Propagating Sausage Mode}
We next determine whether the sausage mode is  propagating along the CS or not by considering the nature of $V_{x}$ at the edge of the CS. We take a vertical slit from $y$ = 20.5 Mm to $y$ = 33.5 Mm at  ($x_{\mathrm{min(B)}}(t)- 0.5$)~Mm, i.e., at the left edge of the CS (as shown by magenta dashed lines in Figure~\ref{label 3}). We consider two cycles of compression and expansion of the cross section of the CS between 1118 s and 1334 s in the absence of plasmoids. We find that the wavy profile of $V_{x}$ at $y$ = 28 Mm at 1130 s clearly propagates outward along the $y$-direction as revealed by the gradual advancement of slowly diminishing perturbations in the form of red, green and magenta curves at 1142, 1154, 1166 s, respectively (see brown arrows in Figure~\ref{label 6}(a)). Another perturbation in $V_{x}$ starts to develop around $y$ = 28 Mm at 1178 s (see the grey curve in Figure~\ref{label 6}(a)) which then propagates outward in the $y$-direction with time (see the advancing wavy profiles as yellow, cyan curves indicated by blue arrows in Figure~\ref{label 6}(b)). 

There is a peak in $V_{x}$ at $y$ = 30 Mm in the blue curve, i.e., at 1214 s, which advances outward as the red curve at 1226 s. A subsequent peak in $V_{x}$ at 1226 s propagates to $y$ = 31 Mm (see the rightmost peak in the green curve in Figure~\ref{label 6}(b)). Later on, another peak at around $y$=26 Mm at 1226 s advances as green, magenta and grey curves at later times as indicated by brown arrows in Figure~\ref{label 6}(b)). Figure~\ref{label 6}(c) also reveals a similar propagation of perturbations in $V_{x}$ with time at later instances between 1262 s and 1334 s, as indicated by brown arrows. Therefore, we infer that the sausage mode is propagating along the CS. Close inspection suggests that once a wavy profile is at the centre of the CS, i.e., around 27-28 Mm, it has a higher amplitude than those closer to the ends of the CS during their advancement in the $y$-direction.

\subsubsection{Leaky Nature of the Sausage Oscillations}
Panels (a), (b) and (c) of Figure~\ref{label 6} suggest that, once the sausage perturbations are propagating outward along the $y$-direction from its centre, their amplitude  diminishes with time. Such an effect may indicate that the energy associated with the sausage mode does not remain trapped totally within the CS. Rather, there may be some leakage of energy from the CS. Therefore, we determine the spatio-temporal variation of the perturbation in $V_{x}$ in the negative $x$-direction, i.e., leftward away from the CS situated roughly at $x$ = 1.5 Mm. This estimate is made at the midpoint of the CS indicated by the lime dashed lines in Figure~\ref{label 3}. We find that $V_{x}$  diminishes in amplitude away from the CS. However, the propagation of additional wavy features is evident in the $x$-direction away from the CS  from 1130 s to 1190 s (see blue, red, green, magenta, grey and yellow curves  as indicated by brown arrows in Figure~\ref{label 6}(d)). 

A similar scenario continues from 1190 s to 1202 s (see yellow and cyan curves in Figure~\ref{label 6}(e)). Between 1214 and 1250 s, such a leaky nature is not observed due to the presence of consecutive sausage modes at the centre of the CS. But after 1250 s, the outward propagation of wavy features in the $x$-direction again becomes evident (see grey, cyan, blue and red curves indicated via brown arrows in Figure~\ref{label 6}(f)). Around 1310 s, another sausage mode propagates towards the centre,  reaching it at 1322 s, which corresponds to an increase in $V_{x}$. But, once it advances in the $y$-direction from the centre, $V_{x}$ starts to decrease again with the advancement of wavy profiles in the $x$-direction, as shown by blue arrows in Figure~\ref{label 6}(f)). Thus, we deduce that the sausage mode is not fully trapped within the CS, but is leaky in the $x$-direction resulting in a gradual decrease in $V_{x}$ at the midpoint of the CS  after the propagation of each sausage mode through the region.

\subsubsection{Comparison of Theoretically Predicted and Wavelet Estimated Period of Sausage Oscillation}
As described in Equation (2), the phase speed of a propagating sausage mode is expected to be the maximum value of the tube speed when the magnetic field is spatially varying within the CS instead of being uniform. For a Harris CS profile, the maximum tube speed is about 0.5$v_{Ae}$, where $v_{Ae}$ is the external Alfv\'en speed. So, in the present simulation, we estimate the average  $v_{Ae}$ and maximum tube speed at each time from 469 s to 1984 s. We find that the average $v_{Ae}$ varies  between 159 and 300 $\mathrm{km\  s^{-1}}$, and the maximum tube speed changes  between 84 and 175 $\mathrm{km\ s^{-1}}$. In other words, the maximum tube speed lies between 0.48 $v_{Ae}$ and 0.66 $v_{Ae}$ with time. As evident from panels (a), (b) and (c) of Figure~\ref{label 6}, the wavelength of the sausage oscillations is almost equal to the instantaneous length of the CS. Therefore, we take $L/(\mathrm{Max} (c_{T}))$ as our estimate for the expected period of sausage oscillations, where $L$ is the length of the CS. We find that expected period varies  between 63 s and 115 s with a temporal average of $\approx $89 s and one standard deviation of $\approx$ 10 s. Therefore, this theoretical prediction is close to the  sausage oscillation period that we observe in the numerical experiment using a wavelet analysis.

\subsection{Oscillatory Stretching and Contraction of the CS along its Axis and its Connection with Sausage Oscillations}
\begin{figure*}
\hspace{-1.5 cm}
\includegraphics[scale=0.275]{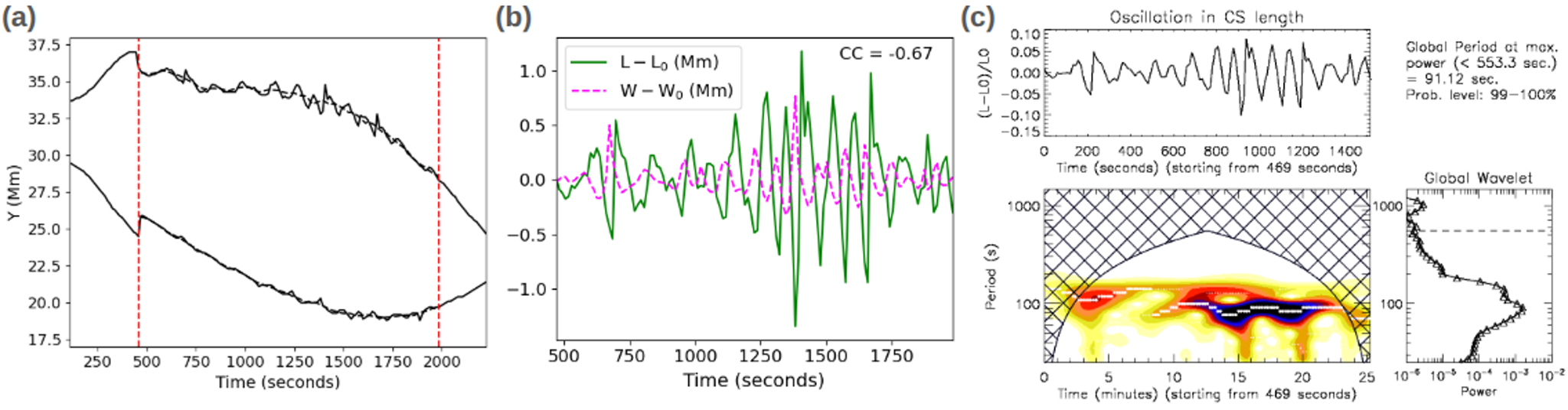}
\caption{\texttt{Oscillatory stretching and contraction of the CS along its axis \textbf{and its correlation with sausage oscillation:}} Panel (a) shows the time-variations in the $y$-locations of the magnetic Y-points  at the top and bottom of the CS. These estimates are carried out at $x$ = $x_\mathrm{min(B)}(t)$ Mm, where $t$ corresponds to the instantaneous times (i.e., along the red dashed lines  shown in the top panel of Figure~\ref{label 2} and the associated animation). Red dashed vertical lines denote the start and end times of the Y-point oscillations. Panel (b) reveals an anti-correlation between $L-L_0$ and $W-W_0$ with cross-correlation coefficient of -0.67 within 469 to 1984 s indicating natural modes of vibration of the CS. Panel (c) exhibits the wavelet estimate of $(L-L_0)/L_0$, where $L_0$ is the distance between the top and bottom Y-points excluding the localized variations in their mean positions at each time. $L$ is basically the distance between top and bottom Y-points. This shows that the CS is oscillating in length with a periodicity of $\approx$ 91 s subjected to a standard deviation of about 6 s.}
\label{label 7}
\end{figure*}

The propagating sausage modes wave-guided by the CS  interact  with the Y-points and cause them to be displaced, and so it is interesting to see whether the CS length varies with the same periodicity as the sausage oscillations. Section 3.3 describes how the locations of the Y-points are tracked in time after $t\approx 108$~s when they form. The top Y-point moves upward while the bottom one moves downward until 469 s (see Figure~\ref{label 7}(a)), after which the top and bottom Y-points move slightly downward and upward, respectively, and then oscillate  until 1984 s (See Figure~\ref{label 7}(a)). Beyond 1984 s, the Y-points converge towards each other as the opposite magnetic flux sources drift apart from each other at the base of the model corona due to numerical diffusion in presence of comparatively weaker converging motion, with the null point at a shifted location. The oscillations of the top and bottom Y-point are not in phase, and so we calculate the  CS length.

Since the CS is elongated and squeezed in the $y$-direction along with its transverse oscillations, we subtract the long-term evolution between 469 and 1984 s marked by vertical red dashed lines (see the dashed black lines overplotted in Figure~\ref{label 7}(a)) to determine [i] the phase relationship between the longitudinal and transverse oscillations of the CS  and, [ii] the periodicity  of the CS length. We subtract the background long-term trend ($L_0$) in the length of the CS  from its actual length ($L=\mathrm{Ypoint_{top}}(t)-\mathrm{Ypoint_{bottom}}(t)$). We find that the oscillations in the CS length ($L-L_0$) and cross-section ($W-W_0$) are anti-correlated with each other with a cross correlation coefficient of -0.67 (see Figure~\ref{label 7}(b)). This means that transverse squeezing of the CS results in an elongation of its length, whereas an expansion in its cross section reduces its length. These oscillations are basically natural modes of vibration of the CS. We then perform a wavelet analysis of $(L-L_0)/L_0$ and find that the periodicity of the oscillation is around $91 \pm 6$ s with a significance level of $99-100~\%$ (see Figure~\ref{label 7}(c)). Therefore, the period of oscillation in the CS length is the same as that of the sausage oscillations. So we next determine the link between these localized natural oscillations of the CS and the large-scale outcomes in the form of arc-shaped wavefronts and high-density separatrix patches. 

\begin{figure*}
\hspace{-1.5 cm}
\includegraphics[scale=0.825]{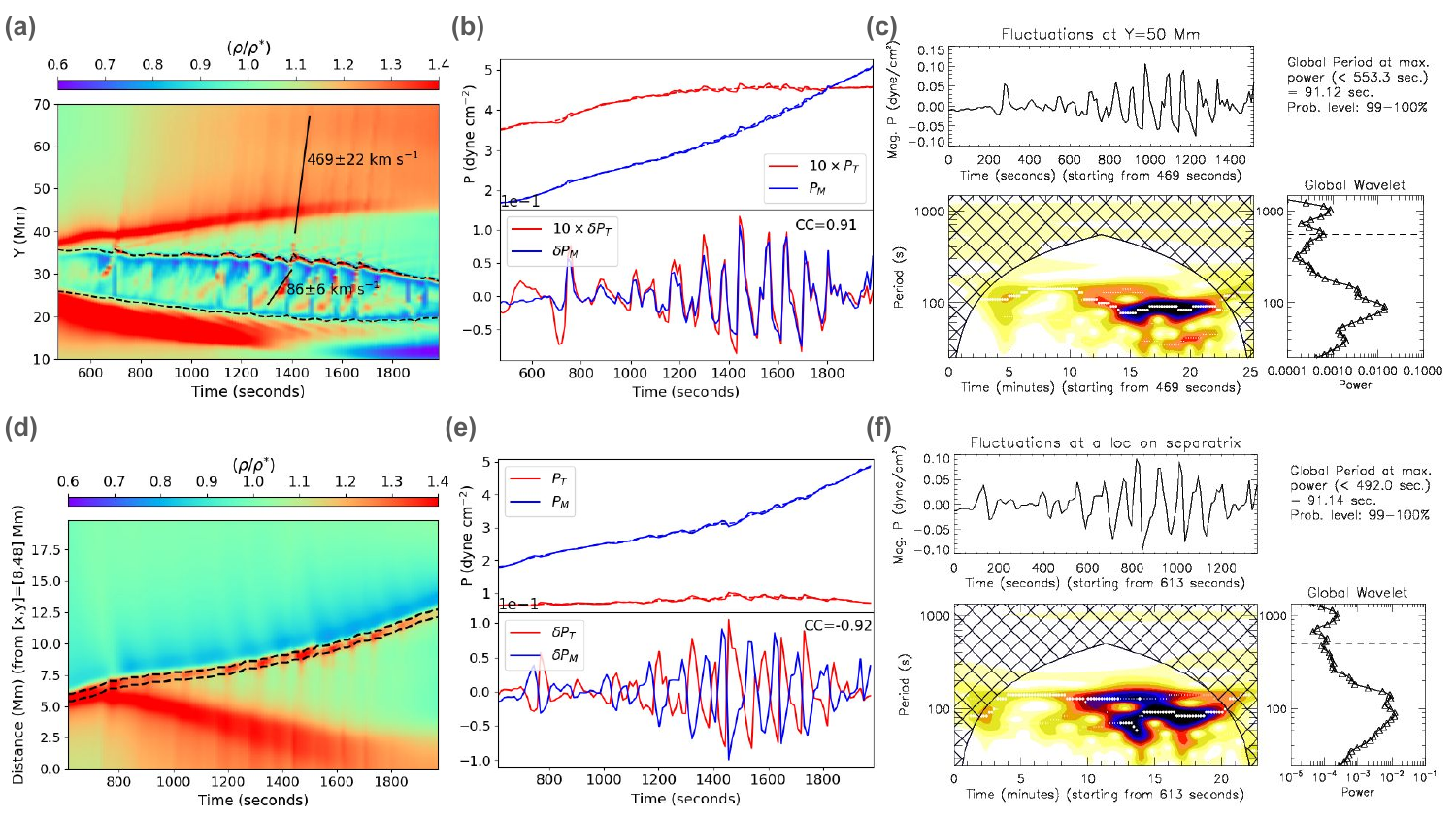}
\caption{\texttt{Identification of fast and slow modes and their connection to CS oscillations:} Panel (a) exhibits a time-distance diagram in density as estimated using a slit extending from $y$ = 10 Mm to $y$ = 70 Mm at $x$ = $x_\mathrm{min(B)}(t)$ Mm (denoted by $\mathrm{S_{f}}$ in Figure~\ref{label 2}). It reveals that the locations of Y-points shown in Figure~\ref{label 3}(a) are associated with the location of density enhancements at the top and bottom of the CS. Rightward slanting ridges indicate outward propagating disturbances along the slit. In (b), the top panel shows the variation in thermal ($P_{T}$) and magnetic pressure ($P_{M}$) with time at $y$ = 50 Mm, while in the bottom panel we subtract the background variation to reveal a strong correlation between the fluctuations in $P_{T}$ and $P_{M}$. This confirms that the propagating disturbances  in panel (a) are fast-mode MHD waves. The periodicity associated with these waves is estimated to be around $91\pm 6$  s (panel (c)). Panel (d) shows the time-distance diagram in density along the slit $\mathrm{S_{s}}$ (see Figure~\ref{label 2}). The density variations (red patches) within the black-dashed curves correspond to disturbances propagating through the intersection points of the slit $\mathrm{S_{s}}$ and the separatrix. In (e), the top panel shows the time-variation in  thermal ($P_{T}$) and magnetic pressure ($P_{M}$)  at those intersection points, while the background-subtracted values in the bottom panel reveal a strong anti-correlation between fluctuations in $P_{T}$ and $P_{M}$. This confirms that the  disturbances propagating along the separatrices are slow-mode MHD waves. Panel (f) reveals the same periodicity of $\approx~91$ s for the slow MHD waves with a standard deviation of about $10$ s.}
\label{label 8}
\end{figure*}
\subsection{Identification of Various Wave Modes Generated by the Top Oscillating Y-Point}
Visual inspection suggests that there may be two different modes of waves generated by the CS oscillations (see the animation attached to Figure \ref{label 2}). Arc-shaped wavefronts are propagating in the large-scale corona in the upward direction from the top Y-point, while highly dense periodic patches are moving along the separatrices. We analyse each of these two morphologically different sets of wavefronts separately below. The former propagate in the ambient corona as large-scale wavefronts across the magnetic field lines, suggesting they are fast magnetoacoustic waves. On the other hand, the latter  exhibit periodic density variations propagating in a  confined way along the magnetic fields of separatrices in a very narrow cone-of-influence, suggesting they are  slow magnetoacoustic waves.

\subsubsection{Identification and Periodicity of Fast Waves}
A straight slit is taken from $y$ = 10 Mm to $y$ = 70 Mm. Since the CS undergoes a rightward movement, we move the slit $\mathrm{S_{f}}$ in Figure~\ref{label 2}(a1)-(a5) by varying its $x$-coordinate according to $x_\mathrm{min(B)}(t)$. The resulting distance-time diagram in density reveals that there are plasma flows within the CS having a propagation speed of around $\mathrm{86~km\ s^{-1}}$ (see Figure~\ref{label 8}(a)). The black dashed curves overplotted on the high-density oscillating regions  indicate a clear fitting of the Y-points (as shown in Figure~\ref{label 8}(a)) with these regions. It is evident that each oscillation corresponds to the generation of a tilted ridge  (see Figure~\ref{label 8}(a)). These tilted ridges correspond to upward-propagating density disturbances. The propagation speed is estimated to be $\mathrm{469 \pm 22~km\ s^{-1}}$ (see Figure~\ref{label 8}(a)). At $y$ = 50 Mm the theoretical fast-mode speed, namely, $(c_{S}^{2}+v_{A}^{2})^{\frac{1}{2}}$  varies  between 400 and 613 $\mathrm{km\ s^{-1}}$ with a temporal average of $494 \pm 61~\mathrm{km\ s^{-1}}$, which is consistent with the above measured propagation speed. The variation in thermal and magnetic pressure is extracted at $y$ = 50 Mm from 469 s to 1984 s. As expected, since $y$ = 50 Mm corresponds to a low-$\beta$ environment, the magnetic pressure is much higher than the thermal pressure  there (see Figure~\ref{label 8}(b)). We subtract the long-term background trend to distinguish the fluctuations in magnetic and thermal pressure which are  associated with the propagating disturbances. A strong cross-correlation with coefficient of 0.91 is found to exist between magnetic and thermal pressure fluctuations (see bottom panel of Figure~\ref{label 8}(b)). This in-phase relation indicates that these propagating disturbances along the slit $\mathrm{S_{f}}$ are fast-mode magnetoacoustic waves. The periodicity associated with these waves is found to be $\approx91~s$ (see Figure~\ref{label 8}(c)), indicating  its association with the  sausage oscillations of the CS.

\subsubsection{Identification and Periodicity of Slow Waves}
Since the separatrices are simultaneously moving rightward and downward, it is difficult to track disturbances propagating along them. A slanted slit $\mathrm{S_{s}}$ shown as a dashed green line in panels (a2)-(a4) of Figure~\ref{label 2} is taken across the right separatrix to extract properties of disturbances passing  along the separatrix. From Figure~\ref{label 8}(d), it is evident that successive high- and low-density patches are present within the region bordered by the two black dashed curves. It is notable that this region  is moving from  around 6 Mm to 12 Mm along the slit, which corresponds to a downward movement of the separatrix in time. The successive high-density patches are associated with the density disturbances propagating through the instantaneous intersection point of the slit and separatrix. We extract the spatially averaged thermal pressure and magnetic pressure variations corresponding to these density patches. The top panel of Figure~\ref{label 8}(e) reveals that the magnetic pressure is higher than the thermal pressure  which indicates the presence of a low-$\beta$ region. The bottom panel of Figure~\ref{label 8}(e) shows a strong anti-correlation between fluctuations in thermal and magnetic pressure, which suggests that these disturbances are propagating as slow magnetoacoustic waves along the separatrix. Figure~\ref{label 8}(f) shows that slow waves  have a periodicity of $\approx$ 91 s, thereby confirming their origin to be associated with the sausage oscillations of the CS. Furthermore, we may compare the theoretical slow-mode speed, namely, $c_S$ with the observed propagation speed, as follows. The sound speed at the point of measurement varies between 175 and 230  $\mathrm{km\ s^{-1}}$ with a temporal average of $209 \pm 15~\mathrm{km\ s^{-1}}$. For comparison, the leading edge of a density patch  advances from $[x1, y1]$ = [15.5, 42.4] Mm to $[x2, y2]$ = [17.9, 43.8] Mm in 12 s (from 1455 s to 1467 s). This gives an estimated propagation speed of about 230 $\mathrm{km\ s^{-1}}$, which is consistent with the theoretical slow-mode speed.

\section{CONCLUSION} \label{sec:conclusion}
The literature suggests that observational signatures such as radio bursts and QPPs are often associated with impulsive reconnection \citep{2000A&A...360..715K,2017A&A...602A.122K}, propagation of large-scale fast waves from flaring regions \citep{2016A&A...594A..96G,2017ApJ...844..149K} and MHD oscillations within the flaring CS \citep{2006A&A...452..343N,2021SoPh..296..185K}. A vast range of periods of QPPs,  from a few milliseconds to several minutes, may indicate that there can be multiple  mechanisms responsible in individual events. Simultaneous theoretical studies suggest that the generation of fast-mode waves may be linked with impulsive reconnection, including the interaction of reconnection outflows or plasmoids with magnetic Y-points or loop-tops, or the coalescence of plasmoids themselves, which give rise to pressure perturbations that excite  various wave modes \citep[e.g.,][and references cited therein]{{2012A&A...546A..49J,2015ApJ...800..111Y,2016ApJ...823..150T,2024ApJ...977..235M}}. \citet{2013A&A...554A.144Y} reported that release times of multiple arc-shaped wavefronts are highly correlated with repetitive radio bursts emitted by accelerated non-thermal electrons, which represent an indirect signature of  energy release by magnetic reconnection. Likewise, the presence of MHD oscillations results in periodic modulation of ongoing reconnection within a CS \citep[e.g.,][and references cited therein]{2006A&A...452..343N,2011ApJ...730L..27N,2018SSRv..214...45M}. All of these studies suggest an inter-linkage between MHD oscillations in a CS, the occurrence of impulsive reconnection, and the generation of fast waves. 

The main result of the present study is that the reconnecting coronal CS formed due to the convergence of opposite-polarity magnetic flux sources undergoes sausage-mode oscillations and drives the propagation of both fast-mode and slow-mode waves into the surrounding atmosphere. The sausage-mode oscillations are identified as natural modes of oscillation of the CS at typically the "tube speed" associated with the CS. The nature of the sausage-mode oscillations is that they are surface modes, which propagate upwards along the CS and leak their energy laterally into the environment. Furthermore, the lateral sausage mode oscillations also lead to an oscillation in the length of the CS as the Y-points at its ends oscillate up and down with the same frequency. Moreover, an anti-correlation between the oscillation of CS length and lateral sausage oscillation confirms that the CS is undergoing a natural mode of oscillation.

The cause of such an oscillation in some experiments would be due to the reflection of fast-mode waves from the outer numerical boundaries of the domain.  We have ruled out such a cause in our experiment, however, since we have adopted  boundary conditions that produce very little reflection from them. In the present study, reconnection has two components: one is  slowly varying, quasi-steady reconnection driven by footpoint motion due to the prescribed converging velocities, but superimposed on that is impulsive bursty reconnection associated with secondary tearing. Therefore, in our numerical experiment, the driver for reported natural oscillations of the CS is the fact that the reconnection is not smooth and laminar but instead is impulsive and bursty due to the occurrence of secondary tearing (or plasmoid instability) \citep{priest86b,loureiro07} in the CS and the propagation of the resulting plasmoids upwards along the CS. In the present case, a key consequence of the natural mode of vibration of the CS and the time-dependent reconnection is that they act as a source for fast-mode waves that propagate outwards from magnetic Y-points.  In addition, they drive slow-mode waves outwards along the separatrices. 

Recently, \citet{2025ApJ...984...36S} suggested that two fundamental plasma processes that are important in in coronal heating, namely, magnetic waves and magnetic reconnection, can coexist and reinforce each other, a process that they termed  `Symbiosis of Waves and Reconnection (SWAR)'. Here we give an example of SWAR by demonstrating that time-dependent reconnection drive surface sausage oscillations in a CS, which can in turn produce fast- and slow-mode waves that propagate into the surrounding corona. Indeed, SWAR is a concept which can occur at diverse spatio-temporal scales and can have profound implications for coronal dynamics \citep{Sri24,2024ApJ...977..235M,2025ApJ...984...36S}. In particular, we conclude  that the observations of short-period QPPs (a few tens of secs) in the large-scale solar corona \citep[e.g.,][]{2019SoPh..294...48R}, may be caused by the dynamics and natural oscillations of a coronal CS . 

In many  observations of coronal oscillations, it is difficult to determine the causes of various wave modes propagating in the large-scale corona. In order to understand the coronal heating aspects of such waves, we need to understand the physics of their source regions. Thus, models such as the one presented here may help to understand the observations of flaring current sheets and the dynamics of the coronal environment. In the present work, we have focused on understanding the  oscillations of a CS, and their role in driving fast and slow magnetoacoustic waves. A detailed parametric study of the energetics of such wave modes and their dependence on a dynamical CS will be explored in the future by combining numerical computations and observations.

Our numerical experiment is two-dimensional, but it is expected that similar features will occur in three dimensions. In particular, CSs will form about null points, separators and quasi-separators and are likely to undergo natural surface sausage oscillations as well as impulsive bursty reconnection, which act as sources for fast-mode waves propagating out in all directions, as well as slow-mode waves propagating along separatrix surfaces. Thus, the reconnection is likely to contribute both directly to coronal heating and indirectly by driving waves that subsequently release their energy into the surrounding atmosphere \citep[e.g.,][]{2006A&A...452..343N,2021A&A...647A..31P,2024MNRAS.534.3133P}. Moreover, since in the present simulation, we only focus on coronal dynamics, we have neglected the effect of gravity, since its inclusion  is not expected to make major changes to the generation of sausage oscillations and  magnetoacoustic waves. However, it would be good in future to include gravity for more accurate quantitative results, especially if modeling CS dynamics and reconnection in lower atmospheric layers.

\section*{Acknowledgments}
The authors are thankful to the anonymous reviewer for constructive suggestions which are helpful to improve the paper. The authors appreciate user-friendly flexible framework of open source MPI-AMRVAC 3.0. S.M. gratefully acknowledges the financial support provided by the Prime Minister's Research Fellowship (PMRF) of India. A.K.S acknowledges the ISRO grant (DS/2B-13012(2)/26/2022-Sec.2) for the support of his scientific research. DP appreciates support through an Australian Research Council Discovery Project (DP210100709). This work was partially accomplished during a visit of S.M. to the University of Newcastle, Australia, which was supported by University of Newcastle and PMRF research grant of IIT (BHU). 

\vspace{5mm}
\software{MPI-AMRVAC, Paraview, Python}

\appendix
\section{Representation of Converging motion and its outcome at the base of the modeled corona}
\begin{figure*}
\hspace{-1.0 cm}
\includegraphics[scale=0.8]{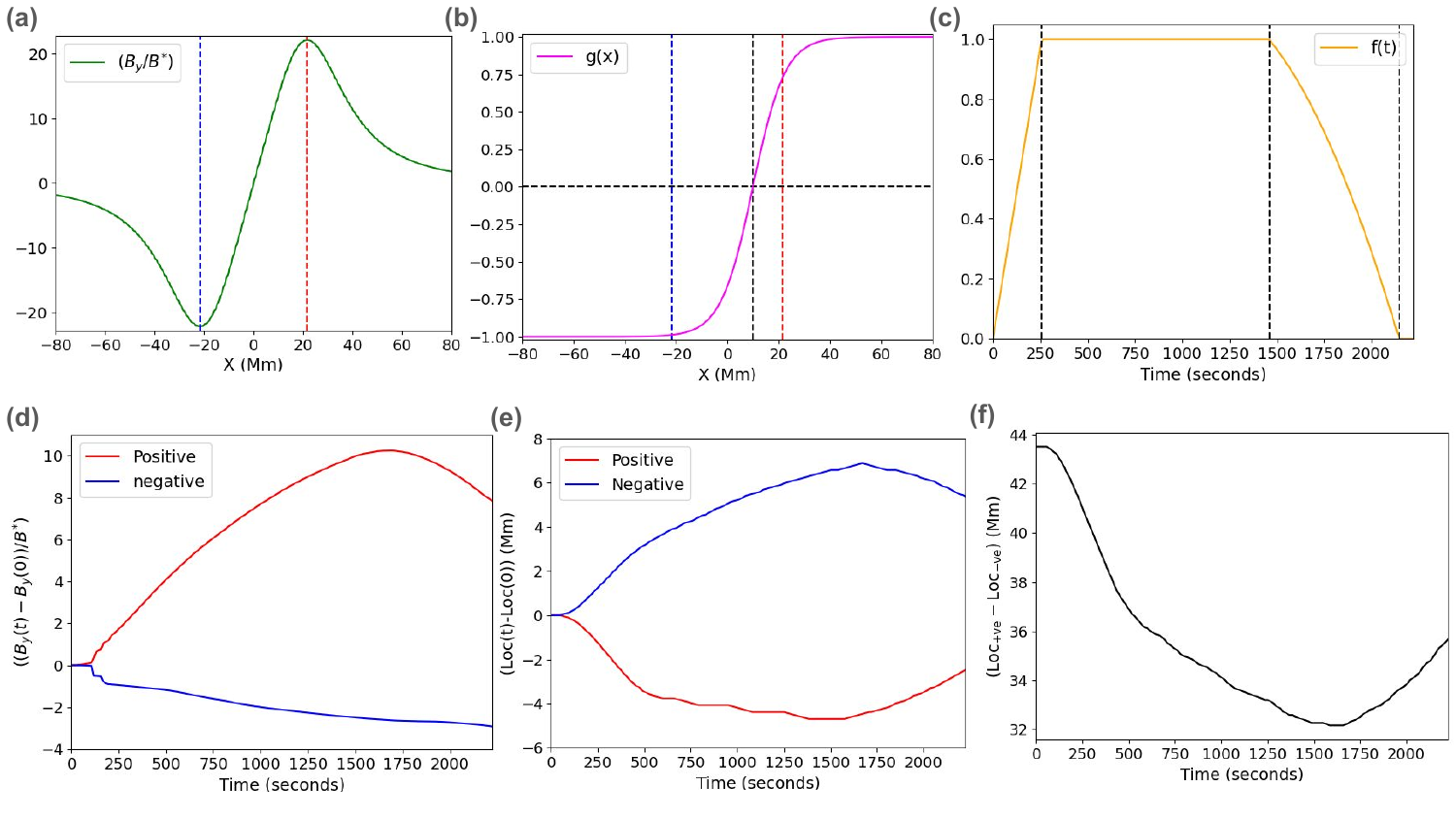}
\caption{\texttt{Visual inspection of the initial configuration at the coronal base and its evolution with time:} Panel (a) shows the spatial variation of the $y$-component of the magnetic field at the bottom boundary at 0 s in dimensionless form. Red and blue dashed lines denote the positions of the peaks of positive and negative magnetic fields at the base of the corona, respectively. Panels (b) and (c) reveal the spatial and temporal variation of the amplitude of the velocity perturbation in terms of $g(x)$ and $f(t)$, respectively (see Equations \ref{eq:g(x} and \ref{eq:f(t)}). Red and blue dashed vertical lines in panel (b) denote the same as in panel (a). This points towards a spatial asymmetry in the imposed perturbation about the location of opposite polarities. Panels (d) and (e) show the change in absolute magnitude of the peak values of magnetic field close to coronal footprints of the opposite polarities and their locations, respectively, with time. Panel (f) exhibits the gradual convergence of the opposite magnetic fields at the base of the corona with time until around 1700 s followed by their drifting from each other due to numerical diffusion in presence of weaker convergence motion.}
\label{label 9}
\end{figure*}

As discussed in the main article, we start with two opposite magnetic polarities placed at a distance of 40 Mm from each other at 20 Mm below the bottom boundary representing the base of the solar corona. 
At the bottom boundary, the peak value of opposite magnetic fields connected to each polarity is $\pm$44 G, and these peaks are separated by 44 Mm at $t=0$ s (see Figure~\ref{label 9}(a)). We impose a spatio-temporally varying velocity perturbation at the bottom boundary. Figure~\ref{label 9}(b) shows the asymmetric nature of this perturbation about the location of the coronal footprints of the opposite polarities (denoted as red and blue vertical lines). Figure~\ref{label 9}(c) represents the variation of the amplitude of velocity perturbation in terms of $f(t)$ (see Equation \ref{eq:f(t)}). Figures~\ref{label 9}(d) and (e) showcase the differences in outcome in terms of the temporal change in signed maximum values and in the positions associated with such peaks in the magnetic field. Since the velocity gradient across the positive polarity is larger (see Figure~\ref{label 9}(b)), the magnetic field becomes more compressed, resulting in a larger increase in its peak value relative to the increase in the peak of the negative polarity (see Figure~\ref{label 9}(d)). Due to the asymmetric driving, the field lines connected to the negative polarity experience a greater displacement than those connected to the positive polarity (see Figure~\ref{label 9}(e)), while the polarity inversion line becomes displaced rightward from its initial position ($x = 0$ Mm). As expected, this displacement is also experienced at the higher height, as seen through the displacement of the collapsed high-$\beta$ environment (see Figure~\ref{label 1} in the main article). Figure~\ref{label 9}(f) reveals the converging motion of magnetic field lines connected to opposite polarities towards each other as witnessed at the base of the modeled corona. In the later stages, after around 1700 s, the flux patches are seen to drift apart due to the effect of numerical diffusion when the converging motion weakens. We have not imposed the source movement directly as was previously done by \citet{2019ApJ...872...32S}. Instead, the plasma velocity is imposed, but the magnetic field slips a little through the plasma when the current density is very large, as is the case in many simulations.
\begin{figure*}
\hspace{-1.0 cm}
\includegraphics[scale=0.8]{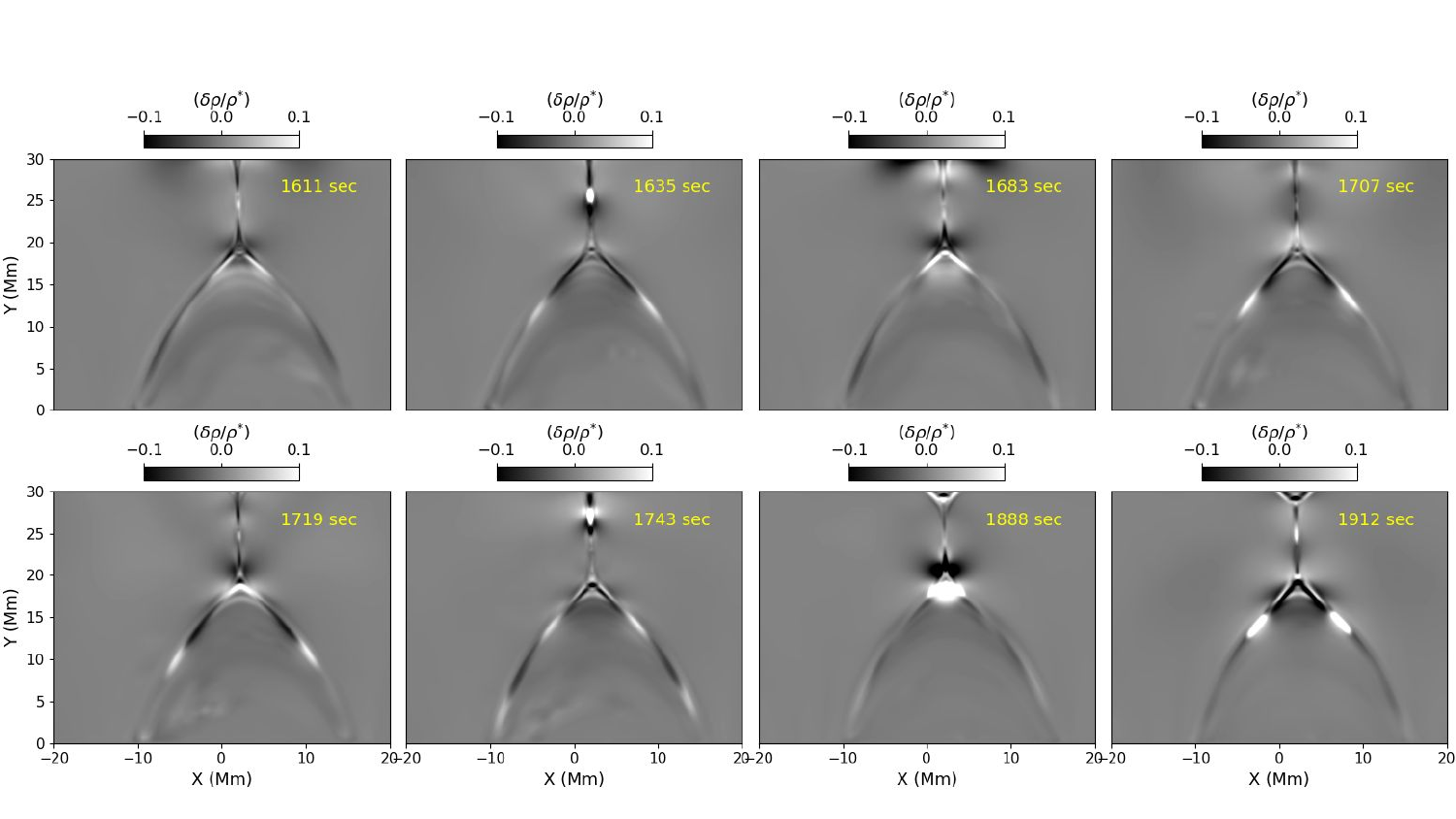}
\caption{\texttt{Visual inspection of the generation of high-density patches at the bottom Y-point and their propagation along the loops:} Running difference maps of density suggest that the bottom Y-point is also perturbed, which further results in the generation of high-density patches propagating along the loops. An animation of real time duration of 4 s covering the dynamics from 469 s to 1984 s is available in the online version. }
\label{label 10}
\end{figure*}
\section{Visual inspection of the generation of slow-mode waves due to an oscillation at the bottom Y-point and its propagation along the loops}
As discussed in the main article, density maps reveal prominent oscillations of the upper Y-point which further result in the generation of fast and slow waves propagating upward and along the upper separatrices, respectively. The lower Y-point exhibits smaller oscillations, which, cannot be tracked convincingly in density maps. However, using running difference images of density, i.e., changes in density distributions at each time from the previous time, we can track such oscillations at the bottom Y-point at multiple times mostly at the later stage of the simulation, i.e., after 1400 s. Such oscillations generate high-density patches which  propagate along the  
the lower separatrices (See Figure~\ref{label 10} and associated animation). Such propagating density enhancements are indicative of slow-mode waves in those loops. However, even after using the running difference method, we cannot find any evidence of fast-mode waves propagating downward from the bottom Y-point which could be due to negligible perturbations in density due to the high Alfv\'en speed at those locations. Even though we find some reflections or upflows propagating upward within the loops, they are mostly limited to low-lying loops (See Figure~\ref{label 10} and associated animation) and cannot reach the bottom Y-point to perturb it and play a significant role in the CS dynamics. 


\end{document}